\documentclass[aps,prl,twocolumn,showpacs,superscriptaddress,preprintnumbers,amsmath,amssymb]{revtex4}
\usepackage{graphicx}
\usepackage{dcolumn}
\usepackage{bm}
\usepackage{color}

\begin{document}

\title{Recoil-induced asymmetry of nondipole molecular frame photoelectron angular distributions in the hard X-ray regime}

\author{\firstname{M.} \surname{Kircher}}
\affiliation{Institut f\"{u}r Kernphysik, J. W. Goethe-Universit\"{a}t, Max-von-Laue-Str.~1, 60438 Frankfurt am Main, Germany}

\author{\firstname{J.} \surname{Rist}}
\affiliation{Institut f\"{u}r Kernphysik, J. W. Goethe-Universit\"{a}t, Max-von-Laue-Str.~1, 60438 Frankfurt am Main, Germany}

\author{\firstname{F.} \surname{Trinter}}
\affiliation{FS-PETRA-S, Deutsches Elektronen-Synchrotron (DESY), Notkestrasse 85, 22607 Hamburg, Germany}
\affiliation{Molecular Physics, Fritz-Haber-Institut der Max-Planck-Gesellschaft, Faradayweg 4, 14195 Berlin, Germany}

\author{\firstname{S.} \surname{Grundmann}}
\affiliation{Institut f\"{u}r Kernphysik, J. W. Goethe-Universit\"{a}t, Max-von-Laue-Str.~1, 60438 Frankfurt am Main, Germany}

\author{\firstname{M.} \surname{Waitz}}
\affiliation{Institut f\"{u}r Kernphysik, J. W. Goethe-Universit\"{a}t, Max-von-Laue-Str.~1, 60438 Frankfurt am Main, Germany}

\author{\firstname{N.} \surname{Melzer}}
\affiliation{Institut f\"{u}r Kernphysik, J. W. Goethe-Universit\"{a}t, Max-von-Laue-Str.~1, 60438 Frankfurt am Main, Germany}

\author{\firstname{I.} \surname{Vela-Perez}}
\affiliation{Institut f\"{u}r Kernphysik, J. W. Goethe-Universit\"{a}t, Max-von-Laue-Str.~1, 60438 Frankfurt am Main, Germany}

\author{\firstname{T.} \surname{Mletzko}}
\affiliation{Institut f\"{u}r Kernphysik, J. W. Goethe-Universit\"{a}t, Max-von-Laue-Str.~1, 60438 Frankfurt am Main, Germany}

\author{\firstname{A.} \surname{Pier}}
\affiliation{Institut f\"{u}r Kernphysik, J. W. Goethe-Universit\"{a}t, Max-von-Laue-Str.~1, 60438 Frankfurt am Main, Germany}

\author{\firstname{N.} \surname{Strenger}}
\affiliation{Institut f\"{u}r Kernphysik, J. W. Goethe-Universit\"{a}t, Max-von-Laue-Str.~1, 60438 Frankfurt am Main, Germany}

\author{\firstname{J.} \surname{Siebert}}
\affiliation{Institut f\"{u}r Kernphysik, J. W. Goethe-Universit\"{a}t, Max-von-Laue-Str.~1, 60438 Frankfurt am Main, Germany}

\author{\firstname{R.} \surname{Janssen}}
\affiliation{Institut f\"{u}r Kernphysik, J. W. Goethe-Universit\"{a}t, Max-von-Laue-Str.~1, 60438 Frankfurt am Main, Germany}

\author{\firstname{L.~Ph.~H.} \surname{Schmidt}}
\affiliation{Institut f\"{u}r Kernphysik, J. W. Goethe-Universit\"{a}t, Max-von-Laue-Str.~1, 60438 Frankfurt am Main, Germany}

\author{\firstname{A.~N.} \surname{Artemyev}}
\affiliation{Institut f\"ur Physik und CINSaT, Universit\"at Kassel, Heinrich-Plett-Str.~40, 34132 Kassel, Germany}

\author{\firstname{M.~S.} \surname{Sch\"{o}ffler}}
\affiliation{Institut f\"{u}r Kernphysik, J. W. Goethe-Universit\"{a}t, Max-von-Laue-Str.~1, 60438 Frankfurt am Main, Germany}

\author{\firstname{T.} \surname{Jahnke}}\email{jahnke@atom.uni-frankfurt.de}
\affiliation{Institut f\"{u}r Kernphysik, J. W. Goethe-Universit\"{a}t, Max-von-Laue-Str.~1, 60438 Frankfurt am Main, Germany}

\author{\firstname{R.} \surname{D\"{o}rner}}
\affiliation{Institut f\"{u}r Kernphysik, J. W. Goethe-Universit\"{a}t, Max-von-Laue-Str.~1, 60438 Frankfurt am Main, Germany}

\author{\firstname{Ph.~V}. \surname{Demekhin}}\email{demekhin@physik.uni-kassel.de}
\affiliation{Institut f\"ur Physik und CINSaT, Universit\"at Kassel, Heinrich-Plett-Str.~40, 34132 Kassel, Germany}

\begin{abstract}
We investigate angular emission distributions of the 1s-photoelectrons of N$_2$ ionized by linearly polarized  synchrotron radiation at $h\nu=40$~keV. As expected, nondipole contributions cause a very strong forward-backward asymmetry in the measured emission distributions. In addition, we observe an unexpected asymmetry with respect to the polarization direction, which depends on the direction of the molecular fragmentation. In particular, photoelectrons are predominantly emitted in the direction of the forward nitrogen atom. This observation cannot be explained via asymmetries introduced by the initial bound and final continuum electronic states of the oriented molecule. The present simulations assign this asymmetry to a novel nontrivial effect of the recoil imposed to the nuclei by the fast photoelectrons and high-energy photons, which results in a propensity for the ions to break up along the axis of the recoil momentum. The results are of particular importance for the interpretation of future experiments at XFELs operating in the few tens of keV regime, where such nondipole and recoil effects will be essential.
\end{abstract}


\maketitle

The presently available X-ray Free Electron Lasers (XFELs) provide a unique opportunity to reinvestigate fundamental questions of light-matter interaction under extreme conditions of ultrashort pulse durations, unprecedented peak intensities, and very short radiation wavelengths \cite{Rev1}. It is, therefore, not a surprise that starting from the early 2000s, the impact of these exceptional properties of the XFEL light on different photoionization processes in the X-ray regime was studied in many experimental and theoretical works. In addition, XFEL facilities  provide unique opportunities for structural imaging of systems which cannot be crystallized and for tracing their time evolution on the femtosecond timescale with angstrom resolution \cite{Rev2}. Improving the spatial resolution of such imaging requires the generation of X-rays with shorter wavelengths. For this reason, several presently operating XFELs, such as the Linac Coherent Light Source (LCLS) at SLAC National Accelerator Laboratory \cite{LCLS} and the European XFEL at DESY \cite{EXFEL}, produce hard X-rays with photon energies of up to approx. 25~keV. One can expect, that these photon energies do not yet represent the upper limit, and even harder X-rays can be generated by XFELs in the near future. This route motivates investigations of the physics of the fundamental properties of light-matter interaction in the photon energy regime of a few tens of keV.

Many interesting phenomena occur when photons with such high energy interact with atoms and molecules \cite{book}. One of the direct consequences of utilizing such short-wavelength radiation, is that the plane wave $e^{i\mathbf{k_\gamma}\cdot \mathbf{r}}$ representing the ionizing field (with $\vert\mathbf{k_\gamma}\vert=\omega/c=2\pi/\lambda$, where $\omega$ is the angular frequency and $\mathbf{k_\gamma}$ the photon momentum vector) cannot  be approximated by unity anymore, and higher order Taylor expansion terms start to significantly contribute beyond the leading electric dipole interaction term  \cite{hxr1,hxr2,hxr3,hxr4}. In almost all existing studies of such \emph{nondipole} interactions, contributions of the higher multipole terms manifest themselves via deviations from dipolar angular emission distributions of photoelectrons \cite{Rev3,Rev4}. Using the velocity gauge and approximating high-energy photoelectrons with momentum $\mathbf{k_\mathrm{e}}$ by a plane wave $e^{i\mathbf{k_\mathrm{e}}\cdot \mathbf{r}}$, one can straightforwardly show that the photoionization matrix element is proportional to the Fourier transform of the initially ionized orbital in the total momentum  space $\mathbf{K=k_e-k_\gamma}$ \cite{JCPeda}, a fact which has recently been used, for example, for orbital imaging \cite{Waitz17}.  As a   consequence, the distribution of the emission probabilities over $\mathbf{k_e}$ exhibits a sizable forward-backward asymmetry, with  considerably more electrons being emitted along the light propagation direction   $\mathbf{k_\gamma}$, i.e., in the forward direction.

Another important effect, which is naturally present in the hard X-ray photoionization regime, is the recoil imparted onto the nuclei by the fast photoelectron \cite{RecPE1}. In molecules, it causes substantial  vibrational \cite{RecPE2,RecPE3,RecPE4} and rotational \cite{RecPE5,RecPE6} excitations, and even phase shifts of photoelectron waves \cite{RecPE7,RecPE8}. Those studies of the recoil by fast photoelectrons are restricted to photon energies of a few to about ten keV. Photons with higher energies carry substantial linear momentum $\mathbf{k_\gamma}$ by themselves and may cause an additional momentum transfer to the nuclei in the forward propagation direction. In the present work, we investigate photoionization of N$_2$ molecules by linearly polarized 40~keV hard X-rays. Thereby, we uncover an unexpected effect of the fast photoelectrons' and high-energy photons' recoil to the ions which manifests in the photoelectron angular emission distribution. Both, the nondipole and the recoil effects,  are huge, and, depending on the molecular orientation in space, their combination breaks the symmetry of the emission distribution with respect to the light polarization axis.

The present experiment has been performed using a cold target recoil ion momentum spectroscopy (COLTRIMS) setup \cite{COLTRIMS1,COLTRIMS2,Jahnke04JESRP} at beam line ID31 of the European Synchrotron Radiation Facility (ESRF) in Grenoble, France. Gaseous N$_2$ is expanded through a 30~$\mu$m nozzle at a backing pressure of 20~bar to create a supersonic gas jet. The gas jet is crossed at a right angle with a 40~keV linearly polarized photon beam yielding a well-defined reaction volume of roughly $0.4\times0.1\times1.0$~mm$^3$. We achieved a photon flux of $8.4\times 10^{14}$~photons/s at an energy resolution of about 1.1\% of $\Delta E/E$ using a pinhole monochromator \cite{pinhole}. A COLTRIMS spectrometer with static electric and magnetic fields was used to guide the charged particles from the reaction volume towards two time- and position-sensitive micro-channel plate detectors with delay line anodes \cite{Jagutzki02}. The active detection area diameter was 80~mm for ions and 120~mm for electrons. The ion and electron arm of the spectrometer consisted of an acceleration region of 13.4 and 27.8~cm length, respectively. The electric field was 51.7~V/cm, and the  magnetic field 20.6~G. This resulted in a detection solid angle of $4\pi$ for the Auger electrons and the Coulomb-exploding N$^+$~+~N$^+$ fragments.

At such high photon energies, the photoelectrons are too energetic to be detected directly. Therefore, only the subsequently emitted Auger electron and the two N$^+$ ions were detected in coincidence and -- by exploiting momentum conservation --  the momentum of the photoelectron is calculated from the momenta of these particles. With the particles momenta, all derived quantities, such as kinetic energies and emission angles, are obtained as well. In particular, the coincident detection of the momenta $\mathbf{p}(\mathrm{N}^+_{L/R})$ of the N$^+$ ions provides access to the two following  quantities:  Firstly, the ion sum momentum $\mathbf{p_{sum}}=\mathbf{p}(\mathrm{N}^+_R)+\mathbf{p}(\mathrm{N}^+_L)$ and, secondly, the relative ion momentum  $\mathbf{p_{rel}}=\mathbf{p}(\mathrm{N}^+_R)-\mathbf{p}(\mathrm{N}^+_L)$. By gating on $\vert\mathbf{p_{sum}}\vert > 30$~a.u., we remove events originating from Compton scattering, because in the case of Compton scattering, the momentum of the emitted electron is not compensated by the parent molecule, and $\mathbf{p_{sum}}$ is small as compared to photoionization (see, e.g., Fig.~1 in Ref.~\cite{Spielberger}).

The momentum difference $\mathbf{p_{rel}}$ provides access to the break-up direction of the N$^+$ fragments. If this break-up happens rapidly after the Auger decay and the molecule does not have time to rotate \cite{Weber01jpb}, the direction of $\mathbf{p_{rel}}$ coincides with the orientation of the molecular axis at the instant of photon absorption as stated by the so-called \emph{axial recoil approximation}. Thereby, the photoelectron angular distribution in the frame of the N$_2$ molecule can be reconstructed as the relative emission angle between the photoelectron momentum $\mathbf{k_e}$ and the difference momentum $\mathbf{p_{rel}}$, obtained from the coincidence measurement \cite{Jahnke2002CON2}.

\begin{figure*}[t]
\includegraphics[width=0.99\linewidth]{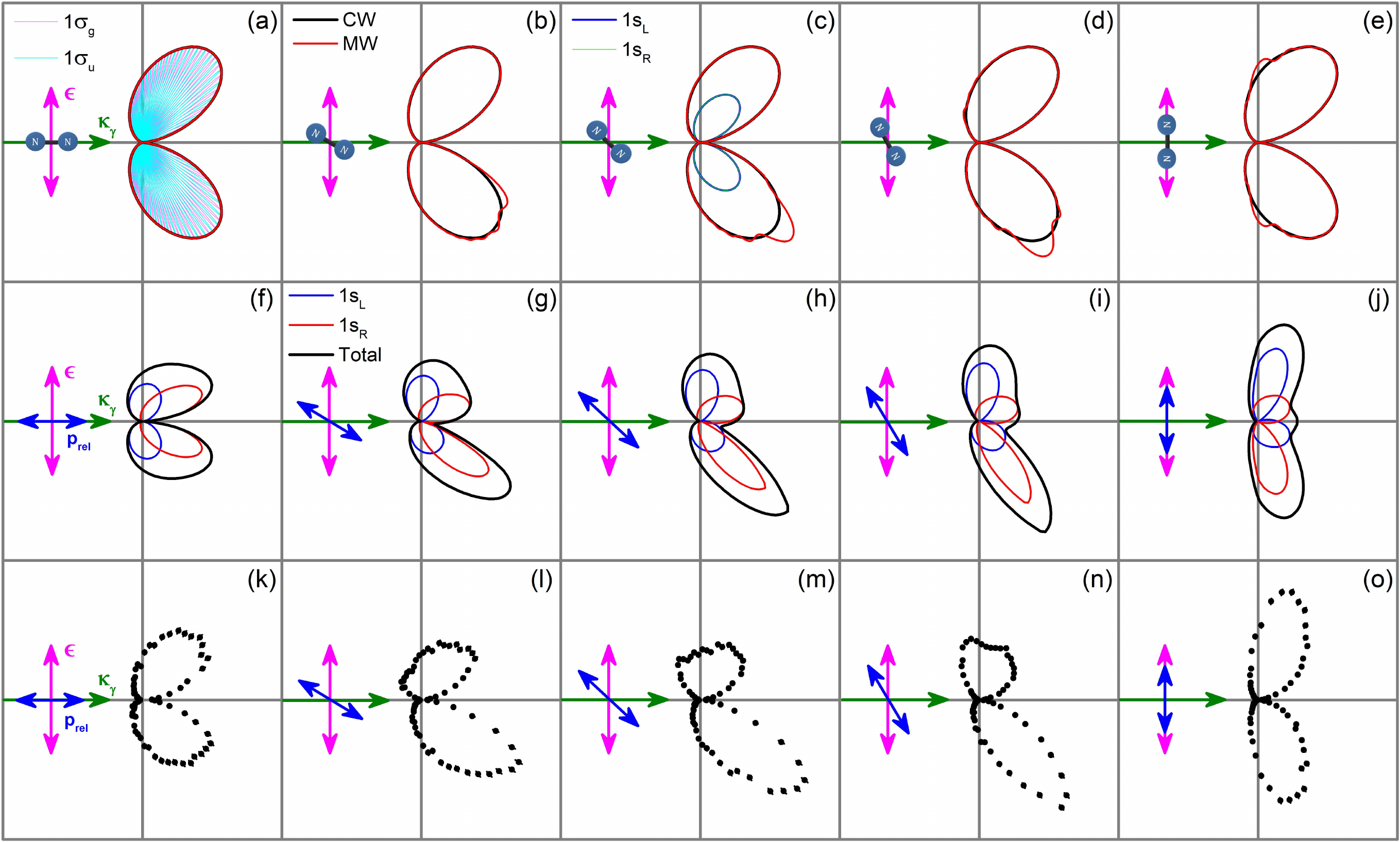}
\caption{Angular distributions of the 1s-photoelectrons of N$_2$ induced by 40~keV synchrotron radiation. The light  propagates from the left to the right, as indicated by the green horizontal arrows, and it is linearly polarized in the vertical direction (see the vertical magenta double-arrows). Upper-row panels (a--e): Present calculations performed using  Coulomb waves (CW) and accurate continuous molecular wave (MW) functions for different spatial orientations of the nitrogen molecule (see the legend in panel (b) and insets in each panel). As an example, the partial CW contributions from the $1\sigma_{g/u}$ and $1s_{R/L}$ orbitals are shown in panels (a) and (c), respectively. Note that the $1s_{R/L}$ contributions in panel (c) are indistinguishable. Middle-row panels (f--j): Results of the present simulations of the combined impact of recoil by the fast photoelectrons and high-energy photons on the partial and total angular distributions [see legend in panel (g)]. The difference of the momenta of the two N$^+$ fragments, $\mathbf{p_{rel}}=\mathbf{p}(\mathrm{N}^+_R)-\mathbf{p}(\mathrm{N}^+_L)$, which in the axial recoil approximation coincides with the molecular axis, is indicated in each panel by the inclined blue double-arrows. Lower-row panels (k--o): Results of the present measurements for different orientations of $\mathbf{p_{rel}}$.}
\label{fig:resultsTE}
\end{figure*}

Panels (k--o) in the lower row of Fig.~\ref{fig:resultsTE} depict the measured photoelectron angular emission distributions for a several selected orientations of $\mathbf{p_{rel}}$ with respect to the photon momentum direction $\mathbf{k_\gamma}$. One can see from Figs.~\ref{fig:resultsTE}(k)--\ref{fig:resultsTE}(o), that the measured angular emission distributions exhibit an enormous forward-backward asymmetry, which is caused by the aforementioned nondipole (retardation) effect. Indeed, almost all photoelectrons are emitted in the forward hemisphere. More surprisingly, for particular orientations of the momentum $\mathbf{p_{rel}}$ in Figs.~\ref{fig:resultsTE}(l)--\ref{fig:resultsTE}(n), the measured distributions exhibit additional huge asymmetries with respect to the direction of the polarization axis of the photons (i.e. up-down asymmetry).

In order to understand these experimental results, the respective photoelectron angular emission distributions were calculated using different approximations within the stationary Single Center (SC) method \cite{SC1,SC2,SC3}. This method has proven to accurately describe angular resolved photoionization and decay spectra of molecules. To rule out any effect of localized initial states of spatially oriented N$_2$ molecules, we first employed the Coulomb wave (CW) approximation. In this approximation, the initial $1\sigma_{g/u}$ or $1s_{R/L}=\frac{1}{\sqrt2}(1\sigma_g \mp 1\sigma_u)$ orbitals were computed in the molecular field, while the final continuum states were approximated by spherical Coulomb waves of energy $\varepsilon \approx 39590$~eV and relativistic momentum $\vert\mathbf{k_e}\vert \approx 55$~a.u. with the effective charge $Z_{eff}=1$. The partial photoionization amplitudes were computed in the velocity gauge including the plane wave $e^{i\mathbf{k_\gamma}\cdot \mathbf{r}}$ in the transition matrix element explicitly by performing the three-dimensional integration numerically. In order to properly describe the underlying nondipole effects, partial electron waves with angular momentum quantum numbers $\ell \le 90$ and $\vert m \vert \le 6$ were included in the calculations.

Results of these calculations are depicted in the upper-row panels (a--e)  of Fig.~\ref{fig:resultsTE} by black solid curves (CW approximation). The  break-ups of the distributions into individual contributions from the  $1\sigma_{g/u}$ and $1s_{R/L}$ initial states are demonstrated in panels (a) and (c), respectively. As one can see, the computed angular distributions exhibit a similarly large forward-backward asymmetry caused by the nondipole contributions, as compared to the experimental distributions in the lower-row. Moreover, the total calculated distributions in the CW approximation are almost equivalent in all panels (a--e). Since the Coulomb waves do neglect the molecular field effects in the final state, one can conclude that the observed up-down asymmetry is not caused by the initial electronic states of the molecule.

In a second step, we included the effect of an asymmetry of the field of a spatially oriented molecular ion on the final photoelectron continuum states. For this purpose, we mixed all partial photoelectron waves by the nonspherical  potential of the molecular ion, by employing the SC method. Results of these calculations are depicted in the upper-row panels (a--e)  of Fig.~\ref{fig:resultsTE} by red solid curves (MW approximation). As one can see, the field of the two nitrogen atoms slightly modifies the angular emission distribution computed in the CW approximation (cf., black and red solid curves in panels (a--e) of Fig.~\ref{fig:resultsTE}). The effect of the molecular field is especially visible in the direction of the forward nitrogen atom. Nevertheless, this weak effect cannot be responsible for the huge asymmetry observed in the experiment.

\begin{figure*}[t]
\includegraphics[width=0.99\linewidth]{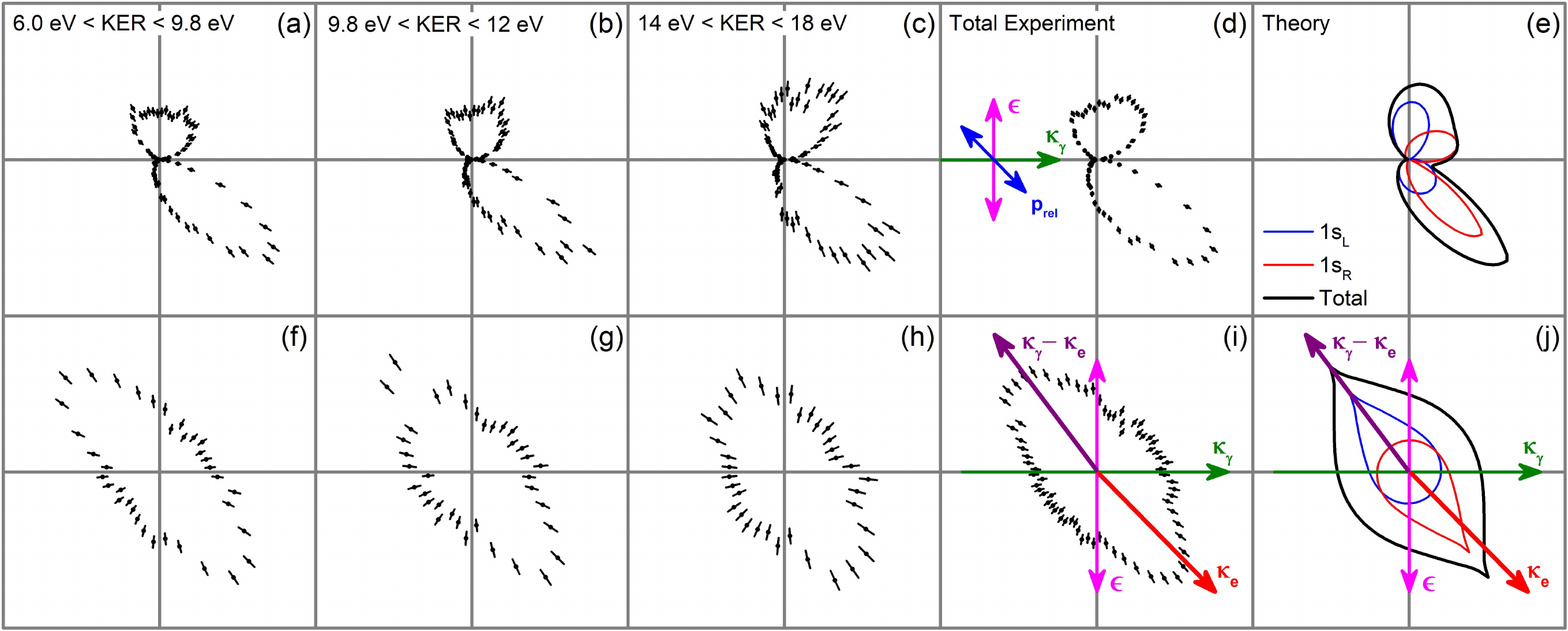}
\caption{Upper-row panels (a--e): Angular distributions of the 1s-photoelectrons of N$_2$, induced by 40~keV synchrotron radiation. Different panels represent measurements for different regions of KER and for all KERs (as indicated at the top of each panel), as well as theoretical partial and total distributions simulating the impact of recoil by the fast photoelectrons and the high-energy photons [see legend in panel (e)]. The experimental geometry is the same as in Fig.~\ref{fig:resultsTE}. The difference of the  momenta of the two N$^+$ fragments, $\mathbf{p_{rel}}=\mathbf{p}(\mathrm{N}^+_R)-\mathbf{p}(\mathrm{N}^+_L)$,  is always kept fixed as indicated by the inclined blue double-arrow in panel (d). Lower-row panels (f--j): Angular distributions of the N$^+$ fragments for a  fixed photoelectron emission angle $\mathbf{k_e}$ [indicated by the red downward inclined arrow in panels (i,j)]. Different panels represent different KER-gating of the experimental data as well as the present theoretical simulations, which are the same as in the upper-row panels. The direction of the combined recoil momentum, $\mathbf{k_\gamma-k_e}$, is indicated in panels (i,j) by the purple upward inclined arrow. The photoelectron emission distributions shown in panels (d,e) are the same as in panels (m,h) of Fig.~\ref{fig:resultsTE}.}
\label{fig:resultsELION}
\end{figure*}

In order to shed light on the origin of the observed asymmetry, we examine the angular distribution of the ionic break-up, i.e.  $\mathbf{p_{rel}}$, for a fixed direction of the photoelectron emission $\mathbf{k_e}$. Within the axial recoil approximation and in the case of the absence of any molecular field effects, such distributions are isotropic. This is confirmed by the fact that the computed total CW distributions in the upper-row panels (a--e) of Fig.~\ref{fig:resultsTE} are independent of the molecular orientation. The presently measured emission distribution of all ions corresponding to a fixed direction of $\mathbf{k_e}$ (indicated by the red downward inclined arrow) is depicted in panel (i) of Fig.~\ref{fig:resultsELION}. As one can see, the measured distribution in Fig.~\ref{fig:resultsELION}(i) is far from being isotropic and strongly aligned along a preferable direction.

A closer inspection of Fig.~\ref{fig:resultsELION}(i) suggests that the ion distribution is almost symmetrically aligned along the direction of the sum of the recoil momentum of the fast photoelectron and the linear momentum of the high-energy photon, $\mathbf{k_\gamma-k_e}$, as indicated by the purple upward inclined arrow. Thus, the momentum provided by high-energy photons with $\vert \mathbf{k_\gamma}\vert=\omega/c \approx 11$~a.u. is not negligible. It looks as if the transfer of this combined momentum enforces the ions to  break up along its direction. If this effect is indeed caused by such momentum transfer, it must depend on the relative strength of this momentum kick to the kinetic energy of the fragments obtained in the Coulomb explosion. Panels (f--h) of Fig.~\ref{fig:resultsELION},  which depict angular distributions of the ions measured for different kinetic energy release (KER) regions, confirm this expectation. We observe that the effect of alignment is large (smaller) for smaller (larger)  KER. A similar dependence is found for the asymmetry in the photoelectron angular emission distributions depicted in panels (a--c) of Fig.~\ref{fig:resultsELION} for the same regions of KER.

In order to support our intuitive explanation, we considered the combined impact of the recoil on the nuclei by the fast photoelectron and the momentum imparted by the high-energy photon theoretically. A full quantum mechanical treatment of the recoil-induced time evolution of a coherent rotational wave packet including the Auger decay and dissociation is a cumbersome task \cite{RecPE6}. However, since the Auger decay is much faster than the recoil-induced molecular rotation (no revival of the rotational wave packet happens within the lifetime of 6.5~fs \cite{N2lifetime}), a simplified classical treatment of the effect successfully explains the present observations. Our classical model includes a recoil-induced rotation of the molecule with a constant velocity before it undergoes Auger decay and consecutively fragments by a Coulomb explosion.

The angular momentum imparted on the molecule depends on the angle between the molecular axis and the recoil momentum. In the present case, it ranges from 0 to about 68$\hbar$, where zero corresponds to the case of a recoil along the molecular axis, and the value of 68$\hbar$ corresponds to the maximal recoil of $\mathbf{k_\gamma-k_e}\approx 66$~a.u. being imparted locally onto one of the N atoms perpendicularly to the molecular axis (here an equilibrium internuclear distance of the ground electronic state of N$_2$ molecule, $r_e=2.074$~a.u. \cite{huber},  was assumed).  This maximal angular momentum transfer yields a rotation of about $19^\circ$ during the Auger decay lifetime of 6.5~fs. As a consequence of a molecular rotation, the theoretical emission distributions computed under the assumption of the axial recoil approximation need to be corrected, depending on the orientation of the combined momentum $\mathbf{k_\gamma-k_e}$ with respect to the molecular axis, the KER value, and the instant of the Auger decay. In addition, the computed distributions need to be corrected for the fact, that the experimentally measured values of $\mathbf{p_{rel}}$ include half of the total momentum $\mathbf{k_\gamma-k_e}$. The respectively corrected distributions were weighted by the exponential distribution of the Auger decay times and integrated over sufficiently large times.

Results of these simulations of the angular emission distribution of the ions are depicted in Fig.~\ref{fig:resultsELION}(j). Calculations were performed for the average value of $\mathrm{KER}=10$~eV and the same orientation of the $\mathbf{k_e}$ vector as in panel (i). The present classical model contains no free parameter. As one can see from Fig.~\ref{fig:resultsELION}(j), the computed distribution of the ions is symmetrically aligned along the $\mathbf{k_\gamma-k_e}$ vector, confirming thereby our  assumption. Moreover, the classical model explains the strong asymmetry of the photoelectron angular emission distributions with respect to the direction of the polarization of the electric field (cf.,  panels (d) and (e) of Fig.~\ref{fig:resultsELION}). Also for different orientations of the $\mathbf{p_{rel}}$ vector, the classically modeled photoelectron emission distributions in the middle panels (f--j) of Fig.~\ref{fig:resultsTE} are in  excellent agreement with the experimental observation (lower-row panels (k--o) of the figure).

In conclusion, we observe huge asymmetries in angular resolved photoemission and fragmentation spectra of N$_2$ molecules ionized by 40~keV synchrotron radiation. Our analysis assigns those observations to the effect caused by the combination of the recoil of the fast photoelectron and the transfer of the linear momentum of the high-energy photons on the nuclei. The recoil-induced molecular rotation creates a propensity for the ions to dissociate along the recoil momentum axis. As a consequence, the photoelectron angular distributions exhibit strong asymmetries with respect to the light polarization axis. The present work provides a show-case example for nontrivial and unexpected effects which can be evoked in the hard X-ray photoionization regime. Knowledge on the presently uncovered effects is a prerequisite for future photoelectron diffraction experiments at XFELs.

\begin{acknowledgements}
The experimental work was supported by the Bundesministerium f\"ur Bildung und Forschung (BMBF) and the Deutsche Forschungsgemeinschaft (DFG) within the Project DO~604/36--2. The theoretical work was supported by the DFG Project DE~2366/1--2. We are grateful for the staff at ESRF for providing the photon beam and thank V.~Honkim\"aki, J. Drenc, H. Isern, and F. Russello from beam line ID31 at ESRF for excellent support during the beam time.
\end{acknowledgements}

\end{document}